\documentclass[sigconf]{acmart}
\usepackage{enumitem}
\usepackage{multirow}
\usepackage{algorithm}
\usepackage{algorithmic}
\AtBeginDocument{%
  }

\setcopyright{cc}
\setcctype[4.0]{by}
\copyrightyear{2025}
\acmYear{2025}
\acmDOI{10.1145/3754598.3754671}
\acmConference[ICPP '25]{54th International Conference on Parallel Processing}{September 08--11,
  2025}{San Diego, CA, USA}
\acmISBN{979-8-4007-2074-1/2025/09}

\begin{document}

\title{LLaMCAT: Optimizing Large Language Model Inference with Cache Arbitration and Throttling}

\author{Zhongchun Zhou}
\authornote{Both authors contributed equally to this research.}
\email{zzhouch@connect.ust.hk}
\affiliation{%
  \institution{The Hong Kong University of Science and Technology}
  \streetaddress{Clear Water Bay}
  \city{Kowloon}
  %\state{}
  \country{Hong Kong}
  }
\orcid{0009-0000-7037-7418}

\author{Chengtao Lai}
\authornotemark[1]
\email{claiaf@connect.ust.hk}
\affiliation{%
  \institution{The Hong Kong University of Science and Technology}
  \streetaddress{Clear Water Bay}
  \city{Kowloon}
  %\state{}
  \country{Hong Kong}
  }
\orcid{0000-0002-9547-9653}

\author{Wei Zhang}
\authornote{Corresponding author}
\email{eeweiz@ust.hk}
\orcid{0000-0002-7622-6714}
\affiliation{%
  \institution{The Hong Kong University of Science and Technology}
  \streetaddress{Clear Water Bay}
  \city{Kowloon}
  %\state{}
  \country{Hong Kong}}

\renewcommand{\shortauthors}{Zhou et al.}

\begin{abstract}
Large Language Models (LLMs) have achieved unprecedented success across various applications, but their substantial memory requirements pose significant challenges to current memory system designs, especially during inference. Our work targets last-level cache (LLC) based architectures, including GPUs (e.g., NVIDIA GPUs) and AI accelerators. We introduce LLaMCAT, a novel approach to optimize the LLC for LLM inference. LLaMCAT combines Miss Status Holding Register (MSHR)- and load balance-aware cache arbitration with thread throttling to address stringent bandwidth demands and minimize cache stalls in KV Cache access. We also propose a hybrid simulation framework integrating analytical models with cycle-level simulators via memory traces, balancing architecture detail and efficiency.

Experiments demonstrate that LLaMCAT achieves an average speedup of 1.26x when the system is mainly bottlenecked by miss handling throughput, while baselines mostly show negative improvements since they are not optimized for this scenario. When the cache size is also limited, our policy achieves a speedup of 1.58x over the unoptimized version, and a 1.26x improvement over the best baseline (dyncta). Overall, LLaMCAT is the first to target LLM decoding-specific MSHR contention, a gap in previous work. It presents a practical solution for accelerating LLM inference on future hardware platforms.
\end{abstract}

\begin{CCSXML}
<ccs2012>
<concept>
<concept_id>10010520.10010521</concept_id>
<concept_desc>Computer systems organization~Architectures</concept_desc>
<concept_significance>300</concept_significance>
</concept>
<concept>
<concept_id>10010147.10010257</concept_id>
<concept_desc>Computing methodologies~Machine learning</concept_desc>
<concept_significance>300</concept_significance>
</concept>
<concept>
<concept_id>10010147.10010341</concept_id>
<concept_desc>Computing methodologies~Modeling and simulation</concept_desc>
<concept_significance>500</concept_significance>
</concept>
</ccs2012>
\end{CCSXML}

\ccsdesc[300]{Computer systems organization~Architectures}
\ccsdesc[300]{Computing methodologies~Machine learning}
\ccsdesc[500]{Computing methodologies~Modeling and simulation}

\keywords{Cache, AI Accelerator, Large Language Model}


\maketitle

\vspace{-1\baselineskip}
\section{Introduction}
% \vspace*{-.7\baselineskip}
% soft: (1) KV Cache, vLLM (2) Decoding
% hard: (1) Cloud + End (2) Cache + MSHR
% TODO cite: GQA, KV Cache, Tensor Core, NPU
% The rapid advancement of large language models (LLMs) has prompted experts in electronic design automation and computer architecture to explore innovative methodologies and chip designs optimized for efficient LLM processing. Despite the diverse applications, LLMs share consistent computational characteristics. They typically adopt a decoder-only transformer architecture and use algorithmic techniques such as the KV Cache and Group-Query Attention (GQA)\cite{GQA}. This uniformity in their algorithmic patterns facilitates the development of domain-specific architectures (DSAs). [TODO]
The advancement of large language models (LLMs) has prompted experts in electronic design automation and computer architecture to explore innovative methodologies and chip designs optimized for efficient LLM processing. Despite the diverse applications, LLMs share consistent computational characteristics. They typically adopt a decoder-only transformer architecture and use algorithmic techniques such as KV Cache and Group-Query Attention (GQA) \cite{GQA}. This uniformity in their algorithmic patterns facilitates the development of domain-specific architectures (DSAs).

LLM inference is primarily bounded by memory bandwidth, making the performance of the memory system a critical factor. This challenge presents an opportunity to design domain-specific and sophisticated architectures that address the distinctive memory and bandwidth requirements of LLMs. However, both consumer- and server-grade chips need a certain degree of generality to adapt to the rapid change of workloads, which limits the utilization of domain-specific accelerators. Consequently, a more practical approach is to integrate the domain-specific design into the existing architecture and share a substantial portion of the system resources with the other components of the chip, such as the NVIDIA Tensor Core and the NPUs used in high-end consumer-grade chips \cite{Intel_AI, xelite} recently. The last-level cache (LLC) is usually one of these shared system resources mentioned above with a significant impact on overall performance.

% This paper follows such a practical approach, focusing on optimizing the LLC for LLM inference through algorithm-hardware co-design.
This paper thus focuses on optimizing the LLC for LLM inference, especially the memory-bound decoding stage.
While traditional cache research has focused heavily on data reuse, LLM workloads exhibit structured data access patterns within dense GEMV or GEMM operators but impose stringent demands on memory-level parallelism and bandwidth. Therefore, our work shifts the focus to optimizing the Miss-Handling Architecture (MHA), which consists of Miss Status Handling Registers (MSHRs) and multiple queues. Efficient MHA utilization is critical, as resource depletion can lead to cache stalls. We address this by improving arbitration policies and introducing a novel throttling mechanism specifically designed for LLM workloads. This approach balances core and memory system interactions, reducing contention and improving overall performance.

Existing AI accelerator performance evaluation toolchains, typically classified into (1) analytical models or (2) cycle-accurate simulators, are often inadequate for LLM-specific research due to the large data sizes of LLMs and hardware architectural complexity. This work thus proposes a framework based on an existing toolchain that combines the strengths of both types to facilitate LLM chip research.

In short, this work makes the following contributions:
\begin{itemize}[leftmargin=0.05in]
    \item We look into the underlying mechanisms of ``cache bandwidth'' and identify the key factors that determine its utilization. We also show that MSHR can be more efficient in capturing temporal locality than cache storage;
    % \item For LLM inference systems, we improve their performance through 3 novel approaches: MSHR- and load balance-aware cache arbitration, together with throttling control;
    \item We design a 2-level dynamic multi-gear throttling policy to have a fine and accurate control over the throttling decisions, in order to relieve contention in the memory subsystem;
    \item To enhance the miss handling throughput of caches, we propose a method to predict the future behavior of memory requests in the arbiter, thereby improving MSHR utilization.
    \item We propose a hybrid framework that integrates an AI accelerator analytical model with a cycle-level simulator by generating memory traces, enabling more accurate performance statistics.
\end{itemize}

\section{Background and Motivation}
% \vspace*{-.5\baselineskip}
Four key factors drive our work:

\begin{enumerate}[leftmargin=0.1in]
    \item LLCs are widely used in LLM inference devices, and they play a crucial role in adapting to algorithmic advances. Although scratchpad memories (SPMs) are generally considered more suitable for DNNs, GPUs and consumer-grade SoCs typically employ LLCs in their designs, due to the diversity of workloads. It is inevitable to optimize LLC for LLM inference in order to benefit from more application scenarios.
    
    \item Previous cache-based research has focused primarily on data reuse, with limited attention to optimizing cache bandwidth utilization and ensuring load balance via arbitration. However, LLM inference is typically bandwidth-bound.
    
    \item The Miss-Handling Architecture (MHA) holds significance in terms of cache bandwidth utilization, although it does not directly enhance the cache hit rate. Contention in the MHA can lead to cache stalls, which can adversely impact overall system performance.
    
    \item Load balancing and appropriate throttling reduce contention and boost performance. Various methods can jointly improve temporal locality and reduce the chance of cache overflow by properly limiting the working set size - a mechanism we reveal in our experiments.
\end{enumerate}
% \vspace*{-1.5\baselineskip}
\subsection{LLM Inference and KV Cache}
% \vspace*{-.5\baselineskip}
% \begin{figure}[!t]
% \centerline{\includegraphics[width=1\linewidth]{fig/LLMInferProcedure.pdf}}
% \caption{KV Cache Mechanism}
% \label{fig:LLMInferProc}
% \end{figure}

\begin{figure*}
\begin{minipage}{0.62\textwidth}
\centering
\includegraphics[width=1\linewidth]{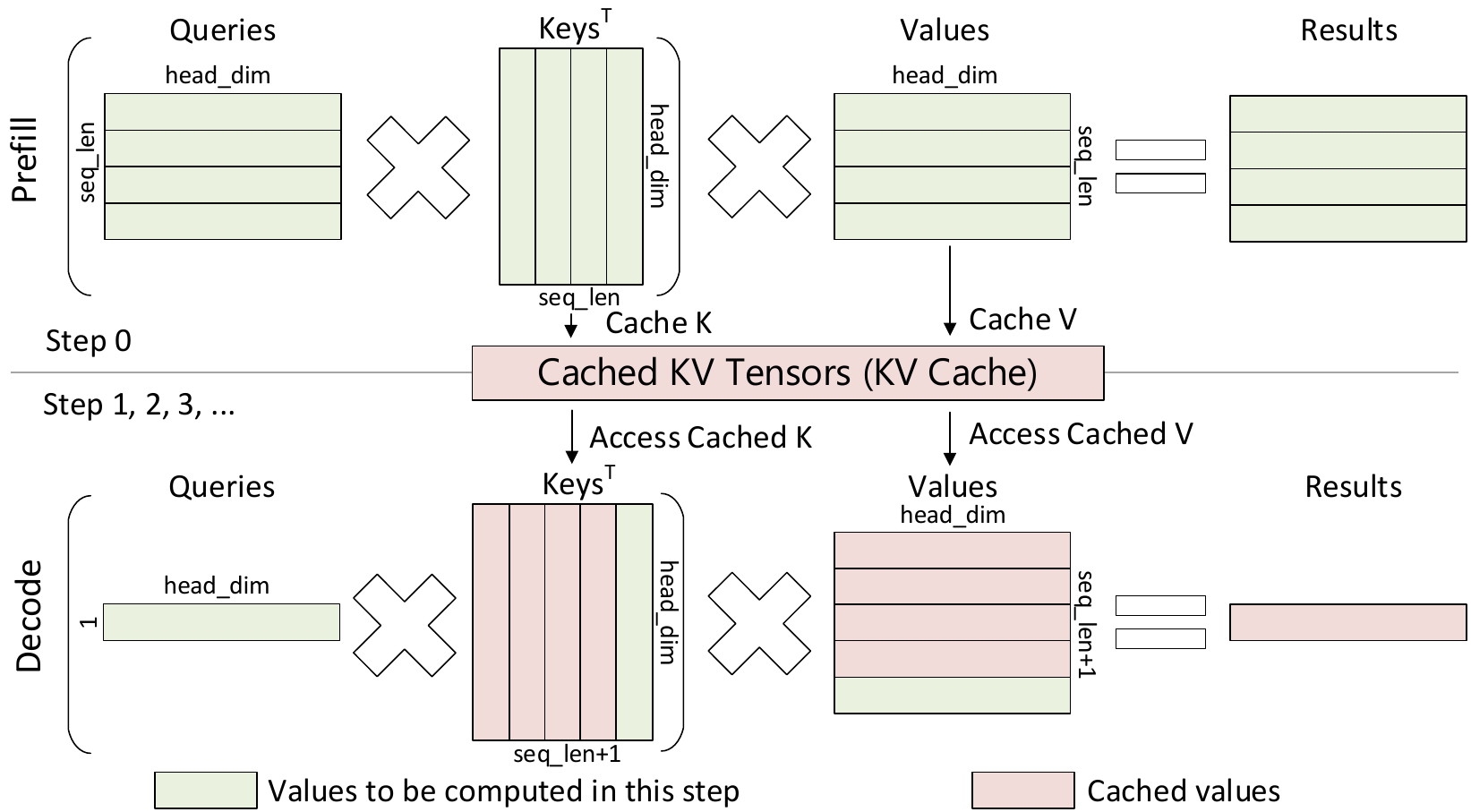}
\vspace*{-1.2\baselineskip}
\caption{KV Cache Mechanism}
\label{fig:LLMInferProc}
\end{minipage}
\hfill  % maximize the space between the minipages
\begin{minipage}{0.37\textwidth}
\centering
\includegraphics[width=.64\linewidth]{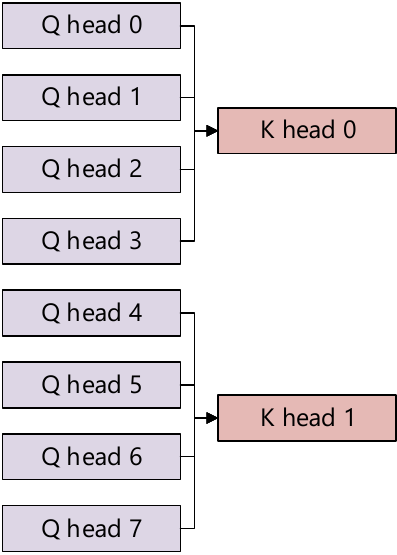}
\vspace*{-.5\baselineskip}
\caption{Group-Query Attention}
\label{fig:GQA}
\end{minipage}
\vspace*{-1.2\baselineskip}
\end{figure*}

Recent advancements have aimed to improve the efficiency of LLM inference, which primarily relies on the auto-regressive, decoder-only transformer architecture. Inference computation can be divided into two stages: Prefill and Decode. The inference process begins with the Prefill stage, which computes the initial Key (K) and Value (V) tensors for the input prompt. In each step afterwards, the model takes in the result of the previous step and generates one token based on \texttt{K} and \texttt{V} tensors of all previous tokens, including the KV tensors of the input tokens computed in Prefill and any new ones computed until the current time step. To avoid repetitive computation, the KV tensors are stored in memory (so called KV Cache); In each time step, new KV tensors are concatenated to existing ones. This process is shown in Fig. \ref{fig:LLMInferProc}.
% During the Prefill stage, the model generates the KV Cache, which represents the context. The KV Cache stores the key and value pairs for the transformer layers, which are reused in subsequent token generation steps. 

% In the Decode stage, the model retrieves the stored KV Cache and generates tokens one by one.
The Decode stage is the most time-consuming and memory access-intensive phase of LLM inference. As demonstrated in \cite{AttAcc}, it dominates the execution time and is highly memory-bound due to frequent memory accesses to the KV Cache and an extremely low computation-memory access ratio (Utilization of GPU compute units can be lower than 1\%). Given that the memory-boundness is primarily attributed to KV Cache access rather than the intermediate results, state-of-the-art operator fusion techniques such as FLAT \cite{flat} and FlashAttention \cite{flashattention3} are not appropriate for this scenario. Efficiently managing these memory accesses is crucial to reducing the overall inference latency.

% \vspace*{-1.5\baselineskip}
\subsection{Group-Query Attention (GQA)}
% \vspace*{-.5\baselineskip}
% \begin{figure}[t]
% \centerline{\includegraphics[width=0.6\linewidth]{fig/GQA.pdf}}
% \caption{Mechanism of Group-Query Attention}
% \label{fig:GQA}
% \end{figure}
To alleviate the memory bottleneck associated with KV Cache, recent research has introduced Group-Query Attention (GQA), which optimizes memory usage by reducing the size of KV Cache. GQA allows multiple query heads to share a single head of KV tensors, effectively reducing the memory footprint without significantly impacting model accuracy, as illustrated in Fig.~\ref{fig:GQA}.

While GQA alleviates memory pressure, it cannot fundamentally resolve the memory bounding issue. The GQA operator employed during the decoding stage remains memory intensive, indicating that our research remains pertinent for such workloads.

Given the adoption of GQA in contemporary LLMs such as Llama3 \cite{llama3}, Gemma2 \cite{gemma2}, and numerous state-of-the-art LLMs, we incorporate GQA-based workloads into our experiments to substantiate the efficacy of our proposed cache optimization techniques for LLM inference.

% \vspace*{-1.2\baselineskip}
\subsection{Optimizing Cache Bandwidth Utilization}
For data-intensive workloads, bandwidth can be more critical than latency. Broadly speaking, cache bandwidth comprises two aspects: the bandwidth of SRAM itself and the capability of handling misses. During cache hits, SRAM bandwidth takes the lead. When cache misses occur, MHA is responsible for storing these outstanding requests and thus prevents the whole cache from being stalled (when cache misses are within a reasonable limit).
% A natural idea is to optimize for a higher hit rate, e.g., by improving the replacement policy or adding a victim cache. However, the hit rate is largely dependent on workloads, especially when the workload is highly regular, like LLM inference, which is mainly composed of GEMV and GEMM. Hence, we must turn to MHA for more improvement.
Compared to optimizing for a higher cache hit rate, the improvement in MHA is much less explored in previous works. The two aspects are also orthogonal to each other: techniques for the latter can be applied together with those for the former. 

As described in Intel® 64 and IA-32 Architectures Optimization Reference Manual~\cite{Intel_Op}, the cache system employs various queues to service various requests from processors and memory controllers, such as the Global Queue (GQ) in the Intel L3 cache system design. Previous research~\cite{POEM, COBRRA} categorizes these queues into two types: request queues, which are responsible for servicing requests from the processor; and response queues, which are responsible for servicing data from the memory controller. These works optimize system performance by managing the request scheduling scheme and priority between the request queues and response queues in conjunction with request bypass to achieve higher system bandwidth. We optimize these arbitration techniques for LLM workloads and finally achieve better performance compared to the baselines.

% \vspace*{-1\baselineskip}
\subsection{Impact of Miss-Handling Architecture}
% \vspace*{-.5\baselineskip}

When MSHRs are exhausted, the cache pipeline will stall, preventing even cache hits from being processed. MSHRs have two key dimensions: \texttt{numEntry}, representing distinct outstanding cache misses, and \texttt{numTarget}, representing requests that can be merged. Cache stalls occur when either dimension is full. It is usually preferred to increase \texttt{numEntry} occupancy while preventing stalls, as \texttt{numEntry} directly influences DRAM bandwidth utilization.

\subsection{Thread Throttling}
% \vspace*{-.6\baselineskip}
\label{throttlebackground}
Thread throttling is effective in mitigating memory system contention by limiting the processor’s front-end throughput. The throttling decision involves three dimensions:
\begin{enumerate}[leftmargin=0.2in]
    \item \textbf{Temporal}: The frequency at which the throttling decision should be modified during execution.
    \item \textbf{Spatial}: The number of cores to which the throttling policy should be applied.
    \item \textbf{Degree}: The extent to which cores should be throttled.
\end{enumerate}

Varying any of these dimensions can have substantial effects on system performance. Previous research \cite{less_is_more} proposes a dynamic throttling logic that applies throttling to all cores. Each core monitors its idle cycles ($C_{idle}$) and memory contention-related stall cycles ($C_{mem}$) independently using its own performance counters. If a core is excessively idle, it relaxes its throttling degree. Conversely, if memory contention is severe (which can be inferred from $C_{mem}$), it increases its throttling degree. In their experiments, the best rate to adjust the throttling degree (i.e., the number of cycles between each throttling degree adjustment) is acquired through parameter sweeping.
However, their method does not handle the \textbf{spatial} dimension, potentially losing an opportunity of finer control. In contrast, our throttling policy allows the number of throttled cores to be dynamically adjusted based on the system contention degree. In addition, the best parameters in their method are obtained by sweeping through a large set of general workloads. If the target workloads are known in advance, e.g., LLM inference in our scenario, the parameters can also be improved.

\vspace{-0.3\baselineskip}
\section{System Architecture}
\begin{figure}[t]
% \vspace*{-0.2\baselineskip}
\centerline{\includegraphics[width=0.68\linewidth]{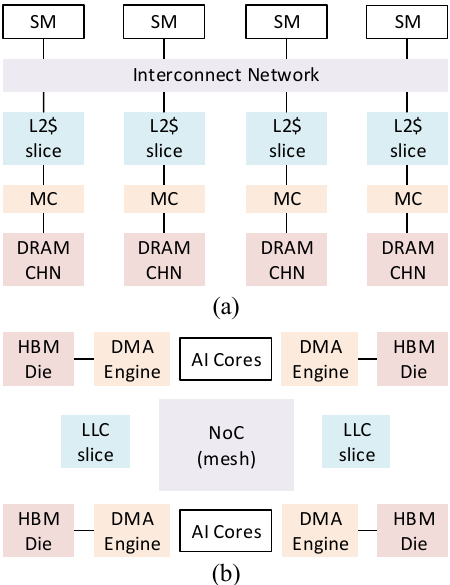}}
\vspace*{-1\baselineskip}
\caption{Example architectures applicable to this work. (a) GPGPU (b) Datacenter-level AI SoC from \cite{Ascend}. SM: Stream Multiprocessor; MC: Memory Controller; CHN: Channel.}
\label{fig:OverallArch}
\vspace{-1.2\baselineskip}
\end{figure}

\vspace{-.1\baselineskip}
\subsection{Overall System Architecture}

\label{overallarch}
% As shared caches are prevalent in today's LLM inference hardware, a large number of architectures fall within the scope of our work. For example, 
Fig. \ref{fig:OverallArch} (a) and (b) show the block diagrams of a typical datacenter-level GPU and an AI SoC with system-level caches. 
% We notice some common characteristics in these architectures: (1) The computing units still have complex internal datapaths and buffers/caches and can hence capture local data reuse to some extent. (2) There are quite a number of such cores (usually tens of them). (3) There is plenty of all-to-all traffic in the interconnect, meaning that each LLC slice is faced with requests from all of the cores. 
Since the dataflow and reuse pattern of NN operators are known at compilation time, both types leverage SPMs in their in-core datapaths. GPU stream multiprocessors (SMs) feature a shared memory that is configurable between SPM and L1 cache, while each DaVinci core of Ascend 910 \cite{Ascend} contains complex datapaths fully composed of SPMs. Despite these differences inside the cores, both include a hardware-managed shared L2 cache. As per FlashAttention-3 \cite{flashattention3}, ``data from global memory gets \underline{transparently} cached into an on-chip L2 cache''. This makes the differences between these cores less critical for predominantly memory-bound LLM inference.
% Although the internal designs of individual cores vary, these differences are less critical for LLM inference, which is predominantly memory-bound. Given that our study focuses on the last-level cache, we simply abstract the upstream compute units as cores in Figure \ref{fig:AssumptionStructure}.
To model the runtime scheduling mechanism in these cores, we adopt an assumption commonly seen in
both types.
% modern GPUs and AI SoCs like \cite{Ascend}.
In each core, we model multiple instruction windows, each of which is assigned a set of instructions once at a time. To align with the terminologies in GPUs, we also call this set of instructions ``thread block''. During execution, to hide memory access latency, when an instruction window is full, the core will automatically switch to another to continue execution. Programmers only have control over the sizes and number of thread blocks, instead of the scheduling algorithms.

\vspace*{-0.2\baselineskip}
\subsection{Baseline LLC Architecture}
We adopt an LLC architecture commonly used in previous works \cite{MRPB, COBRRA}, as shown in Fig. \ref{fig:AssumptionStructure}. In datacenter-level chips mentioned in Section \ref{overallarch}, LLC is usually sliced and physically distributed across the chip. In our setting, we slice the LLC across the cache set dimension, i.e., each slice contains several cache sets. An LLC slice works as follows: (1) a memory request travels through the interconnect and reaches the request queue. For illustrative purposes, we model the request queue as part of the arbiter. (2) The arbiter selects a proper request from the queue and performs cache lookup. If it is a hit, the result is immediately returned to the requesting core (return path not shown); (3) otherwise, it further checks MSHR whether the same address is already pending for DRAM response. If so, the request is merged into that entry. Otherwise, a new entry is opened, and a new memory request is sent to the DRAM. Note this MSHR reservation process will fail if there are not enough resources, and in that case the whole cache pipeline will stall. (4)\&(4') When the request is returned by DRAM, MSHR will first be examined to get the requester(s), and a copy of the data will be directly forwarded to them, while another is pushed into the cache response queue. This ensures that the upstream cores will not wait for the request to travel through the response queue, which may delay their execution. The corresponding MSHR entry is also freed in this phase. (5) When a response dequeues, a bypass manager decides whether to keep the cache line. If not, the data will not be written into cache storage. Since this work is orthogonal to bypass logic design, we do not consider bypassing for fairness and clarity.

% \begin{figure*}[t]
% \begin{minipage}{0.47\textwidth}
% \centering
% \includegraphics[width=1\linewidth]{fig/LLCArch.pdf}

% \caption{System assumption. A slice comprises 1 or more cache sets. Items in red are our design, while others are the baseline. THRTL CTRL: throttle control; cnt: counter for the number of requests served for each core.}
% \label{fig:AssumptionStructure}
% \end{minipage}
% \hfill  % maximize the space between the minipages
% \begin{minipage}{0.51\textwidth}
% \centering
% \includegraphics[width=1\linewidth]{fig/MSHRInfoMergeProcess3.pdf}
% % \vspace{-2\baselineskip}
% \caption{Process of selecting requests from the request queue and updating \texttt{sent\_reqs}. \texttt{spec\_hit\_result}: speculated hit result. In this example, 0x00 and 0xc0 are inferred as cache hits (in grey), so they are not counted when estimating MSHR entries used.}
% \label{fig:MSHRInfoMergeProcess}
% \end{minipage}

% \end{figure*}

% \begin{figure}[t]
% \centerline{\includegraphics[width=0.6\linewidth]{fig/LLCArch.pdf}}
% \caption{System assumption in this work. A slice comprises 1 or more cache sets. Items in red are our design, while others are the baseline. THRTL CTRL: throttle control unit; cnt: counter for the number of requests served for each core.}
% \label{fig:AssumptionStructure}
% \end{figure}

\subsection{Request-response Arbitration}

\label{rra}
\begin{figure}[t]
% \vspace*{-.1\baselineskip}
\centerline{\includegraphics[width=1\linewidth]{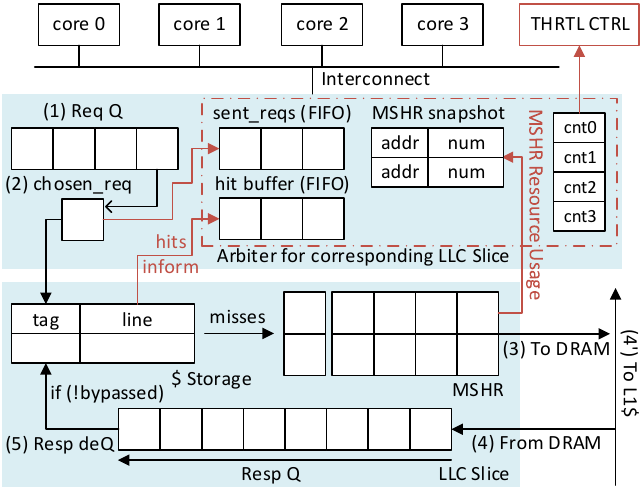}}
\vspace*{-0.7\baselineskip}
\caption{System assumption. Only 1 LLC slice and its corresponding arbiter is shown for simplicity. A slice comprises 1 or more cache sets. Items in red are our design, while others are the baseline. THRTL CTRL: throttling control unit; cnt: counter for the number of requests served for each core.}
\label{fig:AssumptionStructure}
\vspace{-1.3\baselineskip}
\end{figure}

The architectural constraints of the system necessitate careful arbitration between requests and responses, particularly due to the shared nature of certain critical resources. Specifically, as illustrated in Fig. \ref{fig:AssumptionStructure}, both the request path (2) and the response path (5) contend for access to the same cache storage. To manage this contention, a request-response arbitration policy is required. An effective approach, as shown in \cite{COBRRA}, is to prioritize requests over responses, and only when the response queue is full, requests and responses are served in turn. Another approach is to prioritize responses over requests, whenever there is a response to be processed. Experiments reveal that our proposed architectural enhancements yield similar performance gains under both request-response arbitration policies. Without loss of generality, we demonstrate the result of using the latter (response-queue-first) in Section \ref{experimetns}.
% As will be shown in the experiments, our proposed architectural enhancements are agnostic to the specific arbitration policy employed and can yield significant performance improvements under both request-prioritized and response-prioritized arbitration modes. The detailed analysis of the interaction between our modifications and these arbitration schemes is presented in the experiments section.
%================================

\section{CAT: Cache Arbitration and Throttling}

\label{arbdesign}
The main idea of our arbiter design is three-fold: (1) explicitly control MHA to serve the cores on an equivalent basis; (2) throttle the cores whose requests are served more by MHA; (3) make full use of the limited resources in MHA, by using information from LLC and MSHR to guide request selection.

\subsection{Arbiter: Serving Cores Fairly}

\label{balancedmode}
By default, arbiters serve requests on a first-come, first-served basis. Requests served earlier may consume the limited memory resources, potentially causing imbalance between cores. We introduce progress counters to track requests served for each core. The counters are reset at the beginning of each operator execution. Under this mode, the arbiter always picks the request whose requester has the smallest counter value. This policy is named ``B'' (balanced).

% By default, when serving the request queue, the arbiter would treat memory requests on a first-come-first-serve basis. Requests get served earlier may consume the limited memory resources, potentially causing imbalance between cores. To get rid of it, we introduce progress counters that maintain a record of the number of requests served for different cores. These counters are reset at the beginning of each operator execution. Under this mode, the arbiter always picks the request whose requester has the smallest counter value. This policy is hereafter named ``B'' (balanced). [TODO]
% todo: tie-breaking function

\subsection{Dynamic Thread Throttling}

\label{dtt}
To effectively explore the three-dimensional throttling search space (Section \ref{throttlebackground}) for our workload, we highly parameterize the throttling controller design. This design facilitates parameter sweeping across all dimensions and enables dynamic throttling decisions during execution. Besides, we implement a two-level throttling controller: A global multi-gear throttling and an in-core throttling controller, named ``two-level dynamic multi-gear throttling'' (dynmg).

The global dynamic multi-gear throttling controller, illustrated in Table~\ref{table:throttle}, throttles the fastest cores (i.e., those with the largest progress counter values) for load balancing. The gear dynamically changes based on the current system contention degree. Table~\ref{table:con_degree} shows the criteria for classifying contention degree, obtained through parameter sweeping. 
% These parameters are fixed for the edge and cloud device templates in our experiments. During the sampling period, the throttle controller adjusts its gear selection based on the cache’s contention level. 
Algorithm~\ref{alg:dynamic_throttling} shows the multi-gear throttling control logic. High contention triggers a higher gear, while low contention drives the system to a lower gear.

The dynamic policy discussed in the previous paragraph addresses the spatial dimension of throttling decisions, while the \textbf{intra-core throttling decisions still need to be made}. Here we adopt an assumption of throttling the cores similar to the DYNCTA method~\cite{less_is_more}: If a core is throttled, its maximum running thread blocks will be limited. But in our design, we divide each sampling period at the global level into several sub-periods, and decisions on the number of running thread blocks are made in each sub-period. During each sub-period, each core records the number of cycles where all thread blocks wait for memory access ($C_{mem}$). If the cycle count exceeds a threshold, the core reduces the maximum number of running thread blocks ($max\_tb$) by one. Conversely, if the cycle count falls below the threshold, the core increments $max\_tb$ by one. It also records the number of idle cycles ($C_{idle}$). If this cycle count is too high, the core increments $max\_tb$ by one. 

% If and only if a core is selected to be throttled, it will limit its maximum running thread blocks based on the DYNCTA method introduced in work~\cite{less_is_more}. Each sampling period comprises multiple sub-periods in our design. During each sub-period, each core records the number of cycles in which all thread blocks are waiting for memory access (i.e. $C_{mem}$). If the cycle count exceeds a predetermined threshold, it indicates memory contention. In such cases, the core reduces the maximum number of running thread blocks by one. Conversely, if the cycle count falls below the threshold, the core increments the maximum number of running thread blocks by one. It will also record the number of cycles in which the core is idle (i.e. $C_{idle}$). If this cycle is too high, it will increment the maximum number of running thread blocks by one. [TODO]

\textbf{In summary, our in-core throttling controller incorporates two innovative features compared to the original DYNCTA design}: (1) Using DYNCTA as a local logic to throttle individual cores, rather than all cores simultaneously. (2) Introducing two-level sampling periods, where the sampling period in DYNCTA corresponds to the sub-period. Through parameter sweeping, an optimal throttling configuration is identified (Table~\ref{table:throttle_con} and Table \ref{table:incore}). 
% Parameters in Table \ref{table:incore} are fixed for the edge and cloud device templates in our experiments.

% In summary, compared to the original DYNCTA design, \textbf{our in-core throttling controller incorporates two innovative features}: (1) Using DYNCTA as a local logic rather than a global logic, allowing for the throttling of individual cores rather than all cores simultaneously. (2) Introducing two-level sampling periods. The “sampling period” in DYNCTA corresponds to the sub-period employed in our work, as the sampling period in our work is used for the global dynamic multi-gear throttling controller. [TODO]

\begin{algorithm}[t]
\caption{Dynamic Multi-Gear Throttling for Cache Contention}
\label{alg:dynamic_throttling}
% \begin{tablenotes}\footnotesize
% \item $gear$: Current throttling gear
% \item $max\_gear$: Maximum allowable gear
% \item $contention\_status$: Current contention level
% \item $sampling\_period$: Duration of one sampling period
% \end{tablenotes}
\raggedright
$\quad gear$: Current throttling gear \\

$\quad max\_gear$: Maximum allowable gear \\

$\quad contention\_status$: Current contention level \\

$\quad sampling\_period$: Duration of one sampling period \\

\begin{algorithmic}[1]
\STATE Initialize $gear \gets 0$
\WHILE{\textbf{system is running}}
    \STATE Wait for next $sampling\_period$
    \STATE \textbf{Adjust Gear Based on Contention:}
    \IF{$contention\_status ==$ \textbf{High\_Contention}}
        \IF{$gear < max\_gear$}
            \STATE $gear \gets gear + 1$
        \ENDIF
    \ELSIF{$contention\_status ==$ \textbf{Low\_Contention}}
        \IF{$gear > 0$}
            \STATE $gear \gets gear - 1$
        \ENDIF
    \ELSIF{$contention\_status ==$ \textbf{Extreme\_Contention}}
        \IF{$gear \leq max\_gear - 2$}
            \STATE $gear \gets gear + 2$
        \ELSE
            \STATE $gear \gets max\_gear$
        \ENDIF
    \ENDIF
\ENDWHILE
\end{algorithmic}
\end{algorithm}
% \lipsum[3-6]
\setlength{\textfloatsep}{2pt}% Remove \textfloatsep
% \vspace*{-1.6\baselineskip}
% \begin{algorithm}[t]
% \caption{Dynamic Multi-Gear Throttling for Cache Contention}
% \label{alg:dynamic_throttling}
% \begin{tablenotes}\footnotesize
% \item $gear$: Current throttling gear
% \item $m\_max\_gear$: Maximum allowable gear
% \item $contention\_status$: Current contention level
% \item $sampling\_period$: Duration of one sampling period
% \end{tablenotes}
% \begin{algorithmic}[1]
% \STATE Initialize $gear \gets 0$
% \WHILE{\textbf{system is running}}
%     \STATE Wait for next $sampling\_period$
%     \STATE \textbf{Adjust Gear Based on Contention:}
%     \IF{$contention\_status ==$ \textbf{High\_Contention}}
%         \IF{$gear < m\_max\_gear$}
%             \STATE $gear \gets gear + 1$
%         \ENDIF
%     \ELSIF{$contention\_status ==$ \textbf{Low\_Contention}}
%         \IF{$gear > 0$}
%             \STATE $gear \gets gear - 1$
%         \ENDIF
%     \ELSIF{$contention\_status ==$ \textbf{Extreme\_Contention}}
%         \IF{$gear \leq m\_max\_gear - 2$}
%             \STATE $gear \gets gear + 2$
%         \ELSE
%             \STATE $gear \gets m\_max\_gear$
%         \ENDIF
%     \ENDIF
% \ENDWHILE
% \end{algorithmic}
% \end{algorithm}

% \end{table}
% \end{scriptsize}
% \hfill
% \begin{scriptsize}
% \begin{table}[t]

\vspace*{-.6\baselineskip}
\subsection{Arbiter: Select Requests Based on LLC and MSHR Information}

\label{cachemshraided}
The previous two policies aim to maintain balance and slow down the overly-issued cores, while this one opts for efficient utilization of MSHR resources.
% i.e., issuing as many requests as possible before cache stall.
\textbf{Different from request-response arbitration} in Section \ref{rra}, here we focus on determining the order of selecting requests from the request queue to send to LLC.

\vspace*{-.1\baselineskip}
\subsubsection{Motivation and Hardware Units Added}

\label{arbitermotivation}
We notice two facts regarding LLC behavior: (1) cache hits will not lead to a cache stall; (2) cache and MSHR lookup latencies of MSHR hit requests can overlap with DRAM latency since this address is already pending for DRAM response by definition. Suppose that the request arrives later and the previous one has been returned from DRAM, it will become a cache hit. A cache hit would require some extra cycles (hit-latency) before it can be returned to the core. So we conjecture that prioritizing cache hits and MSHR hits can benefit the system by allowing more requests to be sent to the cache before it stalls and reducing average memory access latency.

\begin{scriptsize} 
\begin{table}[t]
  \centering
  \vspace*{-.2\baselineskip}
  \caption{Multi-gear throttling}
  \vspace*{-1\baselineskip}
  \label{table:throttle}
  \begin{tabular}{c|c}
    \hline
    gear & function\\
    \hline
    gear 0 & No throttling \\
    
    gear 1  & throttle 1/8 of the processor cores \\
    gear 2  & throttle 1/4 of the processor cores \\
    gear 3  & throttle 1/2 of the processor cores \\
    gear 4  & throttle 3/4 of the processor cores \\
    \hline
  \end{tabular}
  \vspace*{-.6\baselineskip}
\end{table}
\end{scriptsize}

\begin{scriptsize}
\begin{table}[t]
  \centering
  \caption{Optimal throttling configuration}
  \vspace*{-1\baselineskip}
  \label{table:throttle_con}
  \begin{tabular}{c|c|c}
    \hline
    Domain & parameter & value\\
    \hline
    Temporal & Sampling period & 2000 cycles\\
    Temporal &  Sub-period & 400 cycles\\
    Spatial & Maximum gear & gear 4\\
    Degree & Maximum running thread blocks & dynamic\\
    \hline
  \end{tabular}
  \vspace*{-.6\baselineskip}
\end{table}
\end{scriptsize}

\begin{scriptsize}
\begin{table}[t]
  \centering
  \caption{Cache contention classification}
  \vspace*{-1\baselineskip}
  \label{table:con_degree}
  \scriptsize % Even smaller font size
  \setlength{\tabcolsep}{2pt} % Very small column spacing
\begin{tabular}{c|c}
\hline
Contention degree & Criterion: $t_{cs}$ range \\
\hline
Low & $[0, 0.1)$ \\
Normal & $[0.1, 0.2)$ \\
High & $[0.2, 0.375)$ \\
Extremely High & $[0.375, 1]$ \\
\hline
\multicolumn{2}{l}{$t_{cs}:=$ proportion of cache stall cycles}
\end{tabular}
  \vspace*{-.6\baselineskip}
\end{table}
\end{scriptsize}

\begin{scriptsize}
\begin{table}[t]
  \centering
  \caption{In-core throttling controller parameters}
  \vspace*{-1\baselineskip}
  \label{table:incore}
     \begin{tabular}{c|c}
    \hline
    Parameter & Value (cycles) \\
    \hline
    $C_{idle}$ upper bound & 4 \\
    $C_{mem}$ upper bound & 250 \\
    $C_{mem}$ lower bound & 180 \\
    \hline
\multicolumn{1}{c}{}\
    \end{tabular}
  \vspace*{-.6\baselineskip}
\end{table}
\end{scriptsize}

% We notice two facts regarding cache behavior: (1) cache hits will not lead to a cache stall; (2) cache and MSHR lookup latencies of MSHR hit requests can be overlapped with DRAM latency since this address is already pending for DRAM response. If this request is sent to the cache after the DRAM response comes and thus becomes a cache hit instead of being merged into MSHR, it will take some extra cycles (hit-latency) before the result can be returned to the requesting cores. If these two types of requests are prioritized over the others, the system can potentially benefit from the following two points. (1) More requests can be sent to the cache before it becomes stalled. (2) The cores will experience a shorter average latency for memory requests. [TODO]
% Furthermore, intuitively if request A is a cache hit while request B is an MSHR hit, then $addr_A$ is in a later phase of life cycle than $addr_B$, since $addr_A$ must have once been a cache miss. Prioritizing the former can also potentially help to maintain balance among cores.

We propose a hardware solution to \textbf{identify these types of requests} before actual cache or MSHR lookup (i.e., prediction): adding a \texttt{hit\_buffer} (FIFO) to record recent cache hits and using MSHR information
% an MSHR snapshot (wire from real MSHR, instead of registers) to record the number of occupied targets of each entry}
(\texttt{MSHR\_snapshot}+ \texttt{sent\_reqs}). The \texttt{hit\_buffer} is updated once a new cache hit is determined, while \texttt{MSHR\_snapshot} is passed to the arbiter through a direct wire connection, so it is a real-time summary of the MSHR.

Cache hits can be speculatively recognized by querying the \texttt{hit\_buffer}. For MSHR hits, recall that it takes several cycles (hit-latency) to perform cache lookup, and if a cache miss occurs, it will take some extra cycles (mshr-latency) to lookup the address in MSHR. Thus a newly-sent request causing cache \& MSHR miss will only appear in MSHR and \texttt{MSHR\_snapshot} after hit-latency+mshr-latency. However, during this period, the information about this request is absent from \texttt{MSHR\_snapshot}, without which the arbiter will make suboptimal decisions based on outdated status. To this end, we propose \texttt{sent\_reqs} (FIFO) to keep track of these newly sent requests in aid of \texttt{MSHR\_snapshot}. A request is designed to be dequeued from \texttt{sent\_reqs} after hit-latency+mshr-latency, since it will have been updated in MSHR at that time. When our arbiter queries for speculated MSHR information related to an address, these structures should offer a hint on whether the address is already in the MSHR to help identify MSHR hits.
% \begin{itemize}[leftmargin=0.1in]
%     \item total inflight requests of the cache line, 0 if non-existent (estimated \texttt{numTarget} usage of the MSHR entry). If this number has not reached maximum \texttt{numTarget}, further issuing this address will not cause a stall (i.e., type (b) requests).
%     \item total inflight requests of distinct addresses (estimated \texttt{numEntry} usage). If this number has not reached maximum \texttt{numEntry}, further issuing an address non-existent in MSHR will not cause a stall (i.e., type (c) requests).
% \end{itemize}

\subsubsection{Overall Flow of Request Selection}
\begin{figure}[t]
\vspace*{\baselineskip}
\centerline{\includegraphics[width=1\linewidth]{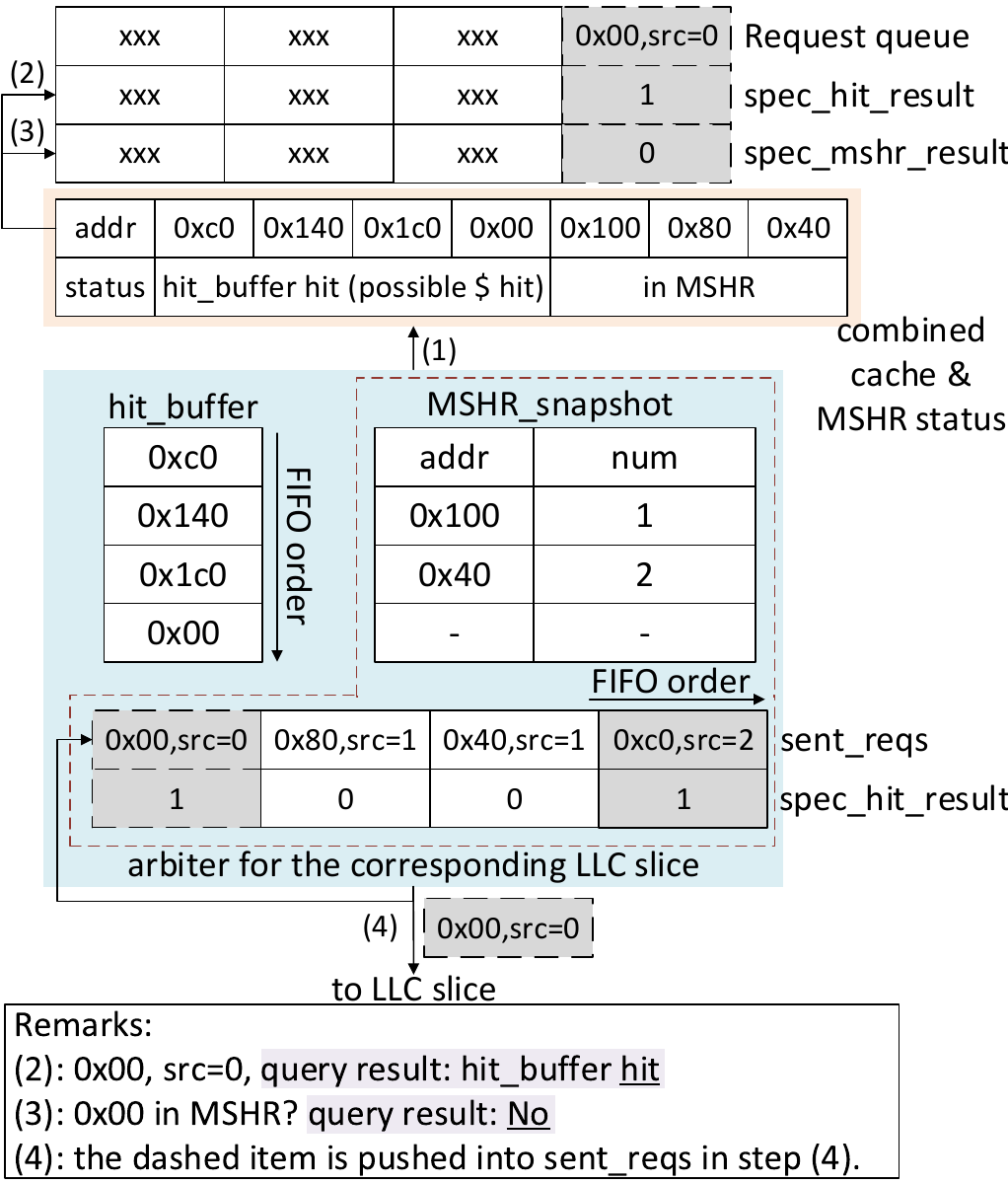}}
\vspace*{-0.6\baselineskip}
\caption{The process of selecting a request from the request queue and updating \texttt{sent\_reqs}. \texttt{spec\_hit\_result}: speculated hit result. In this example, 0x00 and 0xc0 are inferred as cache hits (colored in grey), so they are not counted when estimating MSHR entries used.}
\label{fig:MSHRInfoMergeProcess}
\end{figure}

Please refer to Fig. \ref{fig:MSHRInfoMergeProcess} for this process. When the arbiter begins to select a request from the request queue, it will first combine information from the \texttt{hit\_buffer}, \texttt{MSHR\_snapshot} and \texttt{sent\_reqs} into a list (step 1), where each address is annotated with a status. The type of a request depends on whether it appears in this list. After that, for each element in the request queue, the arbiter looks up the \texttt{hit\_buffer}-related section in the list and generates a \texttt{spec\_hit\_result} bit (step 2). This result indicates the arbiter's speculation on the request being a cache hit. It also looks up the MSHR-related section and checks whether this address is already in MSHR (step 3). When a request is finally chosen (step 4), it is pushed back into \texttt{sent\_reqs} with the \texttt{spec\_hit\_result} bit. When combining \texttt{MSHR\_snapshot} and \texttt{sent\_reqs}, this bit acts as a mask: If it is 1, the corresponding request will not be counted, since MSHR is not involved in cache hits.

\vspace*{-.1\baselineskip}
\subsubsection{Arbitration Policy}
\label{ArbitrationPolicy}
Our policy can be described as rules to rank the priority of requests. It chooses the request with the highest rank to send to the cache:

\begin{itemize}[leftmargin=0.1in]
    \item An inferred cache hit is assigned the highest priority;
    \item An inferred MSHR hit is assigned the second highest;
    \item If the above criteria produce a tie, use a tie-breaking mechanism, which could be either the default request arbitration method or the result of balanced request arbitration (both described in Section \ref{balancedmode}).
\end{itemize}

Later in the experiments, we call the policy with default request arbitration ``MSHR-aware arbitration'' (MA) and the one referring to the balanced result ``Balanced MA'' (BMA).

\section{Simulation Framework}
\label{simframework}
Simulation frameworks are crucial for designing and verifying novel hardware architectures. However, current open-source simulators face limitations when applied to LLM inference. Cycle-accurate simulators of accelerators like SMAUG \cite{smaug} are slow and lack extensibility, especially for large-scale LLM workloads. While they provide detailed intra-core modeling, this level of detail is unnecessary for memory-bound LLM inference \cite{flat}, where the focus should be on global memory performance. Furthermore, SMAUG lacks native support for transformers, requiring operator extensions for LLM simulation. Flexibility of the simulator is another concern when performing architecture exploration. Furthermore, simulators like Accel-Sim \cite{AccelSim}, built upon the GPGPU-Sim framework, present significant hurdles for our research since their design is tightly coupled with detailed replication of specific GPU microarchitectures. These simulators deeply embed GPU-specific assumptions, such as fixed warp sizes, complex warp scheduling logic, and memory coalescing mechanisms based on thread IDs, adding unnecessary complexity and rigidity to our study focused on optimizing shared LLC behavior. Critically, Accel-Sim relies on executing compiled GPU programs (e.g., PTX code), and this dependency on actual kernel code severely hinders efficient dataflow exploration; manually rewriting and compiling GPU kernels for each dataflow variant is far more time-consuming and error-prone compared to studying memory subsystem impacts using traces generated from analytical tools like Timeloop, and is constrained by the GPU programming model itself. Therefore, our hybrid simulation framework, driven by memory traces, more effectively decouples the study of LLC and memory system from the intricacies of specific core implementations, offering greater flexibility and efficiency for our purposes.
% For instance, some assumptions specific to GPUs are deeply embedded into Accel-Sim \cite{AccelSim}. It requires a real GPU program to run, making it difficult to modify the code to drive simulation solely with memory traces, which hinders the exploration of the impact of dataflows on memory subsystem performance.
% \textcolor{blue}{In terms of the accuracy of the memory subsystem, gem5-based simulators fail to provide a detailed modeling of request/response queues and arbitration mechanisms due to their absence in the gem5 classic memory system. The order of serving requests received in the same cycle largely depends on the order in which the API is called, which may cause simulation artifacts.}

On the other hand, AI accelerator analytical models such as Timeloop \cite{timeloop} provide fast emulation, useful for early-stage design exploration, but they oversimplify memory behavior, often assuming a stall-free environment. Most of them also lack cache support, making them unsuitable for cache-related research. Although LCM \cite{LCM} includes a cache model, it reduces cache behavior to mathematical formulas, limiting its ability to validate architectural innovations. To make matters worse, the presence of a cache may break apart contiguous memory accesses to DRAM, leading to crucial DRAM events like row buffer miss. This makes it inappropriate to compute average bandwidth simply with weighted harmonic mean of SRAM and DRAM bandwidth.

The analyses above point out that research on AI accelerators with caches calls for sufficient details of cache and DRAM components such as queues, write, allocation and replacement policies, as well as major DRAM events. It also has to be light-weight enough to simulate a whole LLM operator within reasonable time.
% Our framework addresses these limitations with two key goals: (1) Integrating cycle-accurate memory simulation (cache and DRAM) with simplified intra-core modeling, and (2) Bridging algorithm dataflow, analytical models, and cycle-accurate simulation.
These demands drive us to memory trace-driven, light-weight, cycle-accurate simulators like Ramulator2.0 \cite{ramulator2}.
We keep its DRAM modeling completely unchanged, while dramatically expanding its original \texttt{SimpleO3} frontend, including the following major changes:

\begin{itemize}[leftmargin=0.1in]
    \item Modify each \texttt{SimpleO3Core} to a vector core with a width of 128 elements to align with the head dimension in popular transformer models. Since each thread in a vector group executes commands synchronously, their memory requests can be coalesced by default. This allows us to reduce memory requests by more than an order of magnitude.
    \item Enable a runtime scheduling mechanism that resembles a warp scheduler in GPUs. We model multiple instruction windows in a \texttt{SimpleO3Core}. During runtime, a ``thread block'' is first assigned to each instruction window. When one is full, the scheduler can switch to another to execute.
    \item Introduce a mechanism to send the thread blocks in the trace file of a slow core to a fast core. This is a compensation to the drawback of the original Ramulator2 modeling that each core can only run on its own trace file, which cannot model global scheduling policies that a thread block can be assigned to any cores. \textbf{Without this feature, our baselines would be underestimated} since fast cores have to wait a long time before the slowest core ends.
    % \item Instantiate multiple \texttt{SimpleO3Core}s to model a multi-core scenario;
    \item Add cache policies like allocate-on-fill, write-no-allocate, write-through, while originally Ramulator2 only supports allocate-on-miss, write-allocate, write back. L1\&L2 caches usually have different design considerations.
    \item Split L2 into multiple slices, each of which consists of a number of cache sets.
    \item Model request and response queues and an arbiter that explicitly selects the transaction to feed to L2.
\end{itemize}

\begin{figure}[t]
% \vspace*{-0.2\baselineskip}
\centerline{\includegraphics[width=\linewidth]{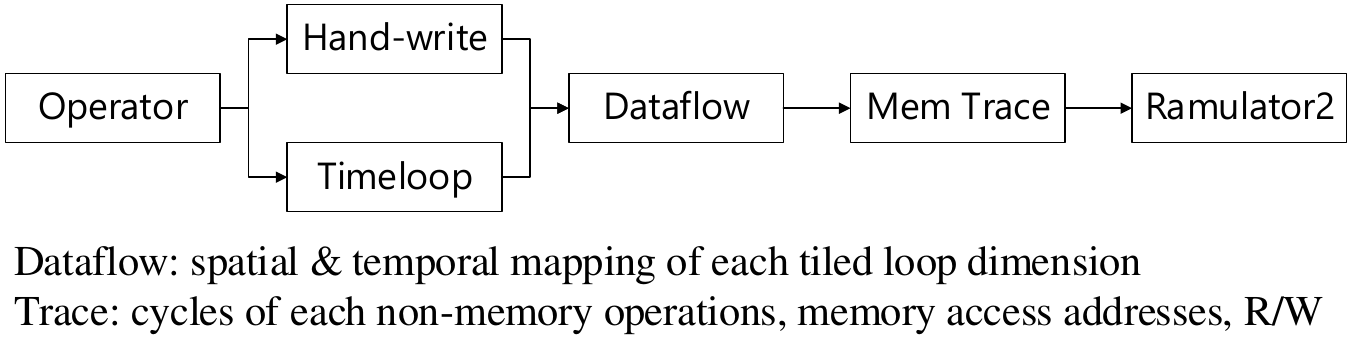}}
\vspace*{-0.7\baselineskip}
\caption{Flowchart of the simulation framework}
\label{fig:flow}
% \vspace*{-1.3\baselineskip}
\end{figure}

To get the memory traces to drive each simulated vector core, we first use Timeloop to generate a mapping of the operator for a given architecture. Since a mapping by definition is a hierarchy of nested loops mapped to either the spatial or temporal domain, it can be translated to memory traces simply by iterating through it. Since the mapping file produced by Timeloop is human-readable, our flow also accepts handwritten mapping dataflows, which is essentially equivalent to adding constraints to Timeloop mapper. We have written a script to seamlessly automate this flow (Fig. \ref{fig:flow}).

% To achieve this, we use Ramulator 2.0 as the DRAM simulator with its \texttt{SimpleO3} front-end for cache modeling and trace-driven multi-core simulation, and we faithfully model our arbiter and related logic in the cache model. We have also developed a trace generation script that supports modern transformer features like GQA and various decoding strategies, including shared prefix and parallel sampling, as shown in Figure~\ref{fig:flow}.

\section{Experiments and Results}
\label{experimetns}
% \vspace*{-.6\baselineskip}
\subsection{Hardware Cost Evaluation}
% \vspace*{-.3\baselineskip}
\label{hardwareeval}
We implement our design in Chisel HDL, and then synthesize the design with Synopsys design compiler and 15nm library\cite{15nm}. The target frequency is 1.96 GHz. All other configurations are as in Table~\ref{table:simconfig}. Synthesis results show that our arbiter and hit buffer have an area of $7312.93\mu m^2$ and $3088.61\mu m^2$, respectively. The design of the arbiter includes the request queue of the cache since they are logically an indivisible unit. The area of the arbiter should be significantly larger than the actual overhead for this reason. 
% The synthesis results are listed in Table~\ref{table:result_s}.
% \begin{scriptsize}
% \begin{table}[t]
%   \centering
%   \caption{Hardware synthesis results}
%   \label{table:result_s}
%   \begin{tabular}{c|c}
%     \hline
%     Module & Area\\
%     \hline
%     Cache Arbiter & 7312.93 ($um^2$) \\
    
%     Hit buffer  & 3088.61 ($um^2$) \\
%     \hline
%   \end{tabular}
% \end{table}
% \end{scriptsize}

% \vspace*{-.7\baselineskip}
\subsection{Experiment Setup}
% \vspace*{-.4\baselineskip}
\subsubsection{Hardware configuration}
A recent industrial trend is the increasing use of consumer-grade devices to run LLMs. Some vendors have announced that their state-of-the-art edge platforms can even run Llama3 70b at a decent generation speed. The simulated hardware setting is listed in Table \ref{table:simconfig}, which is applicable to both high-end consumer-grade devices and cloud chips.
% To show the compatibility of our design to different request-response arbitration policies, we adopt the resp-q-first mode for the edge device and req-q-first for the cloud. [TODOLai]

% To illustrate the key mechanism of our measures, we first use a decoupled setting (Section \ref{efficacyanalysis}) where the L2 miss handling bandwidth is the only bottleneck, \textbf{employing Logit as a demonstrative example}. Then we proceed to a setting closer to reality (Section \ref{largerworkload}) where L2 cache size is also among the bottlenecks, and \textbf{evaluate the whole group-query attention operator.}
Since the policies of concern will display totally different characteristics when the system bottleneck changes, our discussion is divided mainly into two parts according to the sufficiency of L2 cache size (Section \ref{efficacyanalysis} and \ref{largerworkload}).

% For edge devices, we employ the Llama 3 70b model as our workload, while for cloud devices, we utilize the Llama 3.1 405b model due to resource disparities. The comprehensive hardware configurations are detailed in Table~\ref{table:simconfig}.
% \vspace*{-1.4\baselineskip}
\subsubsection{Benchmarks}
Since our main focus is to alleviate the memory bottleneck related to KV cache access, we test our design against the Logit operator ($QK^T$). Computation of this operator is executed across multiple head groups ($H$), head group sizes ($G$), sequence lengths ($L$), and dimensions per head ($D$). The operator sizes are set according to Llama3 70b ($H=8,G=8,D=128$) and Llama3 405b ($H=8,G=16,D=128$). Although Llama3 405b is too large for our current experimental setup, it can be deployed for inference across multiple cards in a pipeline-parallel fashion. Additionally, testing different operator shapes can validate the broad applicability of our improvements.

We add two constraints to the dataflow: (1) assign the fastest axis to each vector core, ensuring cache line access is complete; and (2) map at least 64B of elements in the $L$ dimension to the innermost L1 cache temporal level to avoid false sharing of $AttScore$ among cores. We divide each trace file into contiguous thread blocks for runtime scheduling. Each thread block covers at least 1 cache line of output to avoid false sharing, but larger blocks reduce locality and performance. Empirically, the best performance is achieved when each thread block covers 1-2 output cache lines. Satisfying these constraints results in a hardware-friendly workload that performs well on the unoptimized architecture.

% As for the mapping dataflow, we manually add two constraints: (1) assign the fastest axis (dimension per head) to each vector core as a whole, so that each cache line will not be merely partially accessed and then thrown away instantly; (2) map at least contiguous 64B of elements in the sequence length dimension to the innermost of L1 cache temporal level to avoid false sharing of $AttScore$ among cores. Then we divide each generated trace file into contiguous segments (thread blocks) for runtime scheduling. Each thread block should precisely cover at least 1 cache line of output, otherwise it would again introduce false sharing. But as the thread block grows larger, locality between them would be smaller, and the performance would drop. For the Logit operator, empirically we find the best performance is achieved when each thread block covers 1-2 output cache lines. When the 3 points above are satisfied, we get a hardware-friendly workload that performs relatively well on the unoptimized architecture. [TODO]
% After that we randomly sample from the resulting mappings and select the ones that achieve certain average bandwidths after applying our hardware modifications. Empirically we find 35GB/s for the edge setting and 60GB/s for the cloud are roughly the maximum achievable bandwidths in these benchmarks.

% \vspace*{-1\baselineskip}
\subsubsection{Baselines}
We employ baselines of both throttling (DYNCTA \cite{less_is_more}, LCS~\cite{TB_sche}) and cache arbitration (COBRRA~\cite{COBRRA}) to validate the efficacy of our policies. These baselines are introduced in Section \ref{related_works}.
% To validate the efficacy of throttling and cache arbitration, we employ two categories of baselines for comparison.
For those requiring parameter sweeping, we have also swept under our experiment settings for a fair comparison.

\begin{scriptsize}
\begin{table}[t]
  \centering
  \caption{Simulated System Configurations}
  \vspace*{-.9\baselineskip}
  \label{table:simconfig}
  \scriptsize
\begin{tabular}{p{2cm}p{5cm}}
\hline
Basics & frequency=1.96GHz, 16 cores, 16MB L2 cache, 8 L2 slices \\
\hline
Core & \begin{tabular}[t]{@{}p{5cm}@{}} inst\_window\_depth=128,   num\_inst\_windows=4 \\ 1 core=1 vector unit + private L1 cache, vector-len=128B \end{tabular} \\
\hline
L1 cache & line-size=64B, associativity=8, 64KB, latency=1, alloc-on-fill, streaming, write-no-allocate, write-through \\
\hline
L2 slice & associativity=8, hit-latency=3, data-latency=25, mshr-num-entry=6 (per slice), num-target=8, mshr-latency=5, alloc-on-fill, write-allocate, write-back, req\_q\_size=12, resp\_q\_size=64 \\
\hline
L2 req-resp arbitration & response-queue-first \\
\hline
DRAM & DDR5\_8Gb\_x16, 4 ranks, DDR5-3200, 4 channels \\
\hline
\end{tabular}
  % \vspace*{-.5\baselineskip}
\end{table}
\end{scriptsize}

% \vspace*{-1\baselineskip}
\subsection{Efficacy of Throttling and Arbitration}
\label{efficacyanalysis}
% \vspace*{-.5\baselineskip}
\begin{figure*}[t]
\centering
\centerline{\includegraphics[width=\linewidth]{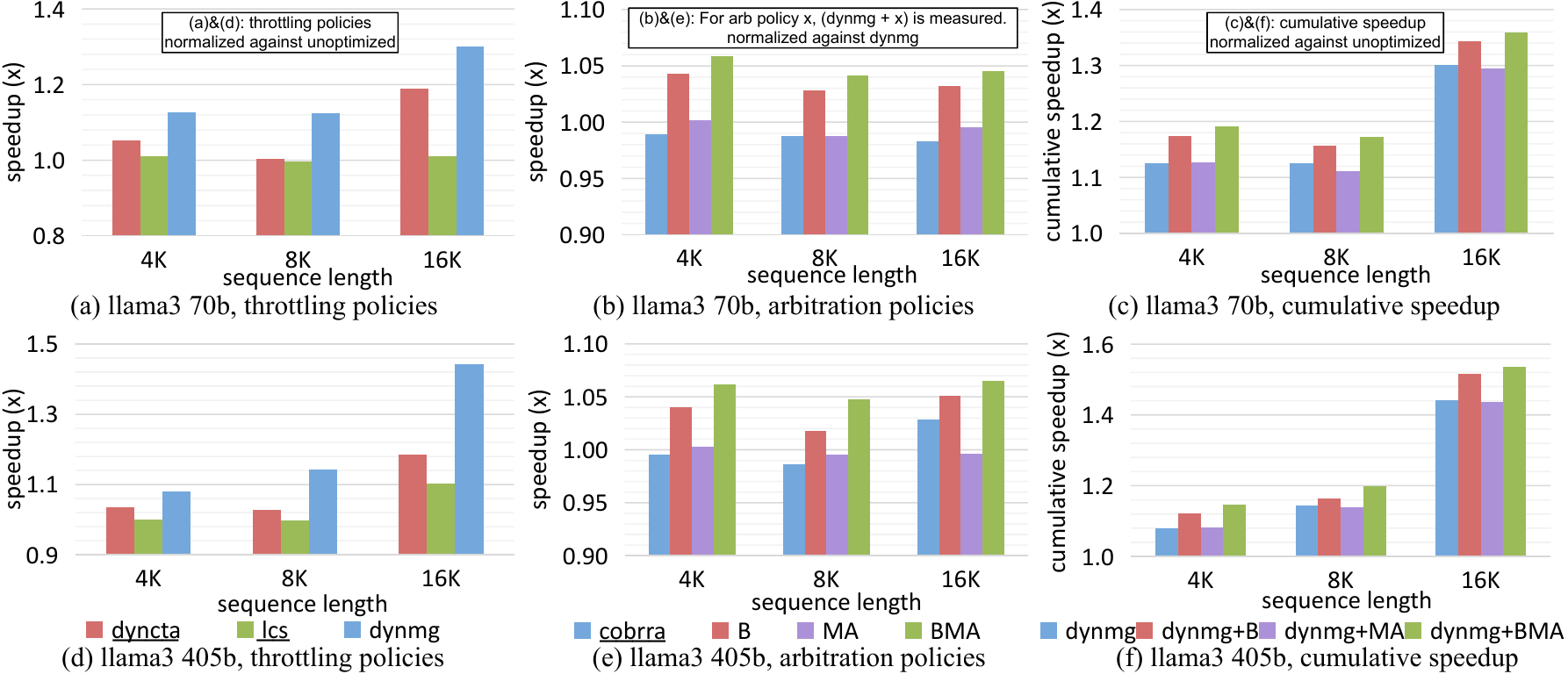}}
\vspace*{-.6\baselineskip}
\caption{Speedups of the Logit operator. All baselines are underlined in the legend. B: balanced; MA: MSHR-aware arbitration. Each policy in (b)\&(e) is aided with ``dynmg''.}
\vspace*{-.6\baselineskip}
\label{fig:arbpolicyresult}
\end{figure*}

\subsubsection{Throttling policies} 
\begin{figure}[!t]
\centerline{\includegraphics[width=1\linewidth]{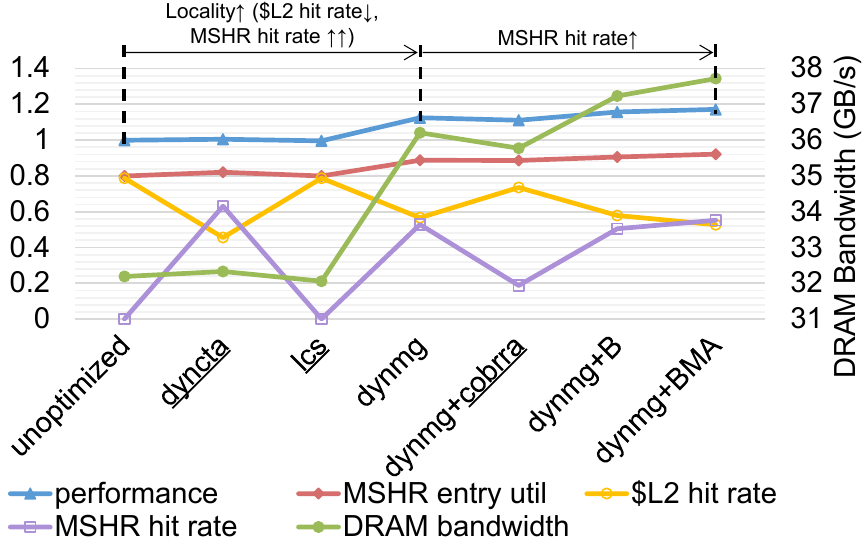}}
\caption{Detailed comparison among policies for llama3 70b 8K benchmark. MSHR entry util: \texttt{numEntry} occupancy throughout execution.}
\label{fig:improvementreason}
\end{figure}

Fig. \ref{fig:arbpolicyresult}(a)\&(d) compare our policy (dynmg) with baseline throttling methods. Speedups are independently normalized against unoptimized versions. Our policy shows 1.08-1.44x speedups (geomean 1.19x), while under most circumstances baseline ``lcs'' does not show meaningful improvements, indicating it is not well-adapted to this scenario. This also applies to baseline ``dyncta'' under shorter sequence lengths. It turns out that these baselines are too conservative: Fig. \ref{fig:improvementreason} will show that their MSHR entry utilization remains almost unchanged compared with the unoptimized version. As Section \ref{largerworkload} will show, these baselines perform better when cache size is also among the bottlenecks. In contrast, our policy can still excavate additional performance gains when workloads are mostly bounded by miss handling throughput.
% Figure \ref{fig:arbpolicyresult}(a)\&(b) compare our policy (dynmg) with baseline throttling methods, where speedups of different hardware configurations and sequence lengths are independently normalized against the execution time of their corresponding unoptimized versions, while Figure \ref{fig:improvementreason} provides a detailed list of other key statistics. Our throttling policy shows a speedup of 1.01-1.13x (geomean 1.09x), while the other baselines mostly exhibit negative improvements, indicating previous throttling mechanisms are not well adapted to this type of scenario. These methods turn out to be overly conservative: Figure \ref{fig:improvementreason} shows that their MSHR entry utilization as well as their MSHR and cache hit rates are all lower than the unoptimized version, demonstrating that the low MSHR entry utilization is not a result of more cache or MSHR hits. This further illustrates that they overestimate memory contention and thus overly throttle the cores to issue fewer requests. As Section \ref{largerworkload} will reveal, these baselines perform much better when the cache size is also among the major performance bottlenecks. In contrast, when faced with workloads predominantly bounded by miss handling throughput, our policy can still excavate additional performance gains. [TODO]

% \vspace*{-1.4\baselineskip}
\subsubsection{Cache Arbitration Policies}
Since arbitration policies are orthogonal to throttling in previous works, here we aid all arbitration policies with the best throttling method (dynmg), to better showcase their joint effect. Fig. \ref{fig:arbpolicyresult}(b)\&(e) compare our versions with baseline ``cobrra''. Speedups are computed by dividing \underline{the performance of} \underline{using each policy with dynmg} by \underline{that of using dynmg only}. We observe that the baseline still causes a performance degradation in most cases. It turns out that the baseline arbitration strategy maintains a stable performance when enabling or disabling throttling. This makes its speedup superseded by
that of the throttling policy.

The results also reveal that ``B'' and ``MA'' need to work together to achieve more improvement. Employing only one of them can also lead to performance degradation. This is because the result of MA sometimes conflicts with the idea of making a fair arbitration among requesters. By returning the result of ``B'' when MA produces a tie, the unfairness is compensated to some extent. Over all cases, our BMA policy produces a speedup of 1.04-1.07x (geomean 1.05x).

% \vspace*{-1.3\baselineskip}
\subsubsection{Core Mechanism of Improvement}
Here we discuss how our policies (unoptimized $\rightarrow$ dynmg $\rightarrow$ dynmg+BMA) produce a speedup in such a scenario. Fig. \ref{fig:improvementreason} lists some other key statistics for analysis. We first notice the number of DRAM accesses does not change dramatically across all policies, but MSHR hit rate keeps improving (annotated for unoptimized, dynmg and dynmg+BMA). In this circumstance, DRAM bandwidth and the throughput of miss handling will take the lead, as shown by the strong correlation among performance, MSHR entry utilization, and average DRAM bandwidth.

Cache and MSHR hit rates, on the other hand, contribute to performance in a more complex mechanism. MSHR hit rate is defined as the number of requests merged to an existing entry divided by the number of cache misses. Cache hits and MSHR hits can convert to each other depending on the timing of requests. In our experiments, they are mostly a result of GQA, since non-GQA operators do not share activation across heads. By properly throttling the cores (unoptimized$\rightarrow$dynmg), the locality of memory requests is enhanced, as shown by the reduced cache hit rate but dramatically increased MSHR hit rate. Hence what we have done at this stage is to \textbf{properly throttle the cores to enhance temporal locality} (and will reduce L2 cache overflow in Section \ref{largerworkload}), \textbf{while still maintaining a reasonable pace to issue memory requests} such that the cores themselves will not become the bottleneck. After that, by adopting cache arbitration policies (dynmg$\rightarrow$dynmg+BMA), MSHR hit rate continues to increase, while cache hit rate keeps decreasing, both in a mild manner. The reason of optimizing towards this direction is that the MSHR lookup latency of an MSHR hit can be overlapped with DRAM latency, while cache lookup latency of a cache hit cannot. In other words, \textbf{MSHR can be more efficient in capturing temporal locality than cache storage}. As shown in Fig. \ref{fig:arbpolicyresult}(c)\&(f), \underline{our final policy (dynmg+BMA) delivers a speedup} \underline{of 1.15-1.54x (geomean 1.26x)}.

\subsection{When Workloads Become Larger}
% \vspace*{-.5\baselineskip}
\label{largerworkload}

% \begin{figure*}[t]
% \begin{minipage}{0.51\textwidth}
% \centering
% \centerline{\includegraphics[width=1\linewidth]{fig/ImprovementReason2.pdf}}
% % \vspace*{-1\baselineskip}
% \caption{Detailed comparison among policies for the 8K benchmark. MSHR entry util: \texttt{numEntry} occupancy throughout execution.}
% \label{fig:improvementreason}
% \end{minipage}
% \hfill  % maximize the space between the minipages
% \begin{minipage}{0.47\textwidth}
% \centering
% \centerline{\includegraphics[width=1\linewidth]{fig/LargerWorkload5.pdf}}
% % \vspace*{-.7\baselineskip}
% \caption{Throttling and arbitration policies running with 16K sequence.}
% \label{fig:largerworkload}

% \end{minipage}
% \end{figure*}

\begin{figure}[t]
\centerline{\includegraphics[width=1\linewidth]{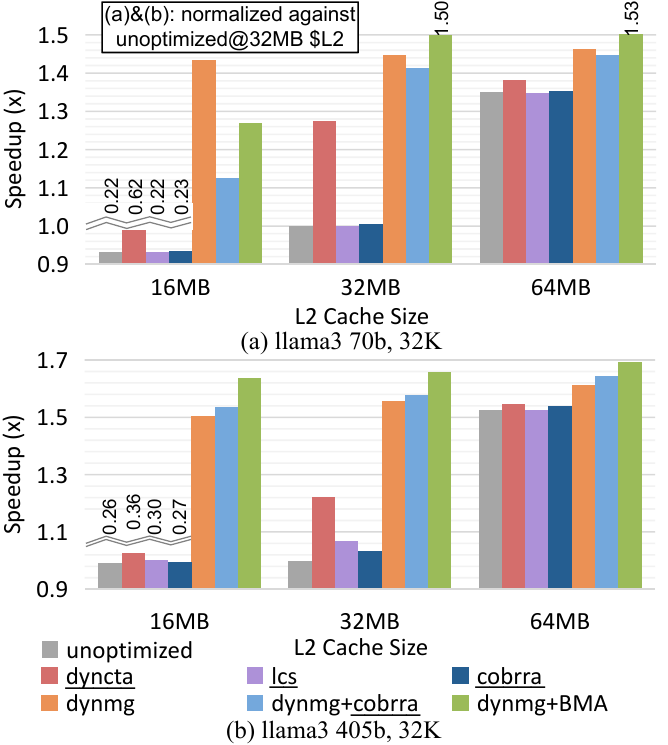}}
\vspace*{-1\baselineskip}
\caption{Throttling and arbitration policies running with 32K sequence.}
\label{fig:largerworkload}
\end{figure}

Real-world applications may require a longer context length. 
% so that miss handling throughput is not the only major bottleneck. 
Longer sequences induce larger working sets, thus imposing more pressure on the cache size. In this section, we will demonstrate that our policy is more resistant to performance degradation caused by limited L2 cache sizes. To achieve this goal, we set the cache size around the original while running 32K sequences for both benchmarks (only 16K in the previous section).

Fig. \ref{fig:largerworkload} shows the execution time of the operator using multiple throttling and arbitration policies. We notice as the cache size varies, performance of the unoptimized version changes dramatically, while that of our policy nearly saturates at 16MB. The unoptimized version’s demand for larger caches stems from its runtime scheduling mechanism: Each instruction window is assigned a thread block, and a core will switch to another instruction window upon stalls. Since the assigned thread blocks may span a wide range, constraining instruction window switching is effective in reducing the working set.

% We notice the unoptimized version continues to speedup as the cache size increases, while all baselines and our methods saturate at a 16MB cache. This illustrates that all of them have some resistance to such performance degradation caused by limited cache sizes. The reason that the unoptimized version requires a larger cache is the runtime scheduling mechanism. Remember that each instruction window is assigned a thread block at a time, and when an instruction window gets stuck, the core switches to another to continue execution. The thread blocks assigned to the instruction windows in the cores can span a large range. Thus limiting the total working set is equivalent to reducing the range spanned by these assigned thread blocks. Thus throttling can be effective in reducing the working set size. [TODO]
Among baselines, dyncta achieves notable speedups over the unoptimized version at 16MB and 32MB but its superiority diminishes at 64MB, suggesting it is better suited for cache-constrained scenarios. Similarly, cobrra outperforms the unoptimized version at 32MB, but its gains diminish at 16MB and 64MB. The dynmg+cobrra combination also outperforms dynmg for the llama3 405b model. In contrast, our final policy, dynmg+BMA, delivers the highest speedup across all cases since it is also capable of efficiently utilizing MSHR resources to enhance miss handling throughput, with the only exception of llama3 70b @16MB. In summary, under the 32MB configuration, our final policy achieves a speedup of 1.50-1.66x (geomean 1.58x) over the unoptimized version, and a 1.18-1.35x (geomean 1.26x) improvement over the best baseline (dyncta).

\section{Related Works}
\label{related_works}
\subsection{On-Chip Memory Optimization for AI Accelerators}
Previous research on optimizing on-chip buffers for AI workloads focuses on either scratchpad memory (SPM) management or cache management. Software frameworks like Pin-or-Fuse \cite{PorF} and OnSRAM \cite{onsram} enhance SPM utilization, but cache-based systems cannot directly benefit from these. LCM \cite{LCM} optimizes cache usage for LLM inference by improving data reuse within the cache but overlooks load balancing and bandwidth optimization. Additionally, it does not consider advanced LLM inference techniques such as Group-Query Attention and KV Cache optimization.

% \vspace*{-.2\baselineskip}
\subsection{Last-Level Cache Arbitration}
While cache replacement policies have been extensively studied to optimize data reuse, few works address bandwidth optimization and scheduling among cores in the LLC. COBRRA \cite{COBRRA} demonstrates that maximizing cache hit rate does not guarantee optimal performance, highlighting limitations in previous research. REAL \cite{REAL} schedules shared LLC bandwidth based on the LLC instruction density in each core's Reorder Buffer. COBRRA combines cache bypassing and arbitration to reduce cache contention and improve system performance. POEM~\cite{POEM} aggressively bypasses cache writeback to reduce cache contention and also optimizes the endurance of NVM cache. However, these works focus on traditional CPU applications and do not address cache stalls caused by MSHR reservation failures.

% \vspace*{-0.6\baselineskip}
\subsection{Miss Handling Architecture}
Most prior works target the utilization of MSHR \texttt{numEntry}. For example, \cite{mshrAwareScheduler} dynamically adjusts GPU warp scheduling: employing higher thread parallelism when few MSHR entries are reserved and lower parallelism when occupancy is high. DL-MSHR \cite{DMSHR} mitigates stalls from insufficient \texttt{numTarget}s by redesigning MSHRs into a list structure, but this incurs significant architectural changes, hardware costs, and additional latency. MRPB \cite{MRPB} addresses \texttt{numEntry} insufficiency through request prioritization but does not tackle limitations related to \texttt{numTarget}.

% \vspace*{-0.7\baselineskip}
\subsection{Thread Throttling}
Our experiments show that throttling alone does not fully optimize performance, but is more effective when combined with other strategies. We focus on integrating throttling with an MSHR-aware cache arbitration policy, not discussed in previous works. DYNCTA \cite{less_is_more} dynamically monitors runtime statistics to adjust the number of active thread blocks based on memory system contention and core busyness. In contrast, LCS \cite{TB_sche} improves throttling by observing the execution of the first thread block to calculate the optimal number of thread blocks without dynamic tuning.

\section{Conclusion}
We introduce LLaMCAT, a comprehensive approach to optimizing the LLC for LLM inference. By synergistically combining MSHR- and load balance-aware cache arbitration with dynamic thread throttling, it effectively addresses the stringent bandwidth demands and improves overall performance.
It achieves a geometric mean speedup of 1.26x when the system is mainly bottlenecked by miss handling throughput, while the baselines mostly cause a performance degradation since they do not optimize for this scenario. When cache size is also a bottleneck, it reaches a speedup of 1.58x, 1.26x over the best baseline. This proves that our enhancements are crucial for future hardware platforms aiming to efficiently handle the memory-bound nature of LLMs.

Moreover, we presented a hybrid simulation framework that bridges the gap between AI accelerator analytical models and cycle-level simulators. This framework, which generates memory traces from dataflow produced by analytical models, offers a flexible and efficient means to evaluate architectural innovations for LLMs, addressing the limitations of existing simulation tools.

\begin{acks}
This work is partially
funded by Hong Kong RGC GRF 16214123 and AI Chip Center for Emerging Smart Systems (ACCESS).
\end{acks}

\bibliographystyle{ACM-Reference-Format}
\bibliography{refs}

% \appendix









\end{document}